\documentclass{moriond}

\def\be{\begin{equation}}
\def\ee{\end{equation}}
\def\bea{\begin{eqnarray}}
\def\eea{\end{eqnarray}}

\newcommand{\Photo}{}
\usepackage{overpic}

\begin{document}
\vspace*{4cm}
\title{ALPs searches at BESIII}

\author{ Peicheng Jiang\footnote{On behalf of the BESIII Collaboration.} }

\address{School of Physics and State Key Laboratory of Nuclear Physics and Technology\\
Peking University, Beijing 100871, China}

\maketitle\abstracts{
A search for an axion-like particle with 2.7 billion $\psi(3686)$ events collected by the BESIII detector is presented. No significant signal is observed, and the upper limits on the branching fraction of $J/\psi\rightarrow\gamma a$ and the ALP-photon coupling constant $g_{a\gamma\gamma}$ are set at the 95\% confidence level in the mass range of $0.165\leq m_a\leq 2.84$ GeV/$c^2$. The limits on $\mathcal{B}(J/\psi\rightarrow\gamma a)$ range from $8.3\times10^{-8}$ to $1.8\times10^{-6}$ over the search region, and the constraints on the ALP-photon coupling are the most stringent to date for $0.165\leq m_a\leq 1.468$ GeV/$c^2$.}

\section{Introduction}

Axion-like particles (ALPs) are pseudo-Goldstone bosons arising from some spontaneously broken global symmetry, addressing the strong CP \cite{axion_1,axion_2} or hierarchy problems \cite{hierarchy}. ALPs could appear in theories beyond the Standard Model (SM) \cite{Higgs}, and also provide a portal connecting SM particles to the dark sectors \cite{dark_1}.
In the most common scenarios, the ALP $a$ predominantly couples to photons with a coupling constant $g_{a\gamma\gamma}$, assuming the branching fraction of $a$ decaying to photons is 100\%. 
As a generalization of QCD axions, ALPs have arbitrary masses and couplings which are bounded by experiments. In the O(GeV) region, the limits of ALP couplings mainly come from electron-positron colliders.

\begin{figure}[hp]
  \centering
  \mbox
  {
  \begin{overpic}[width=0.4\textwidth]{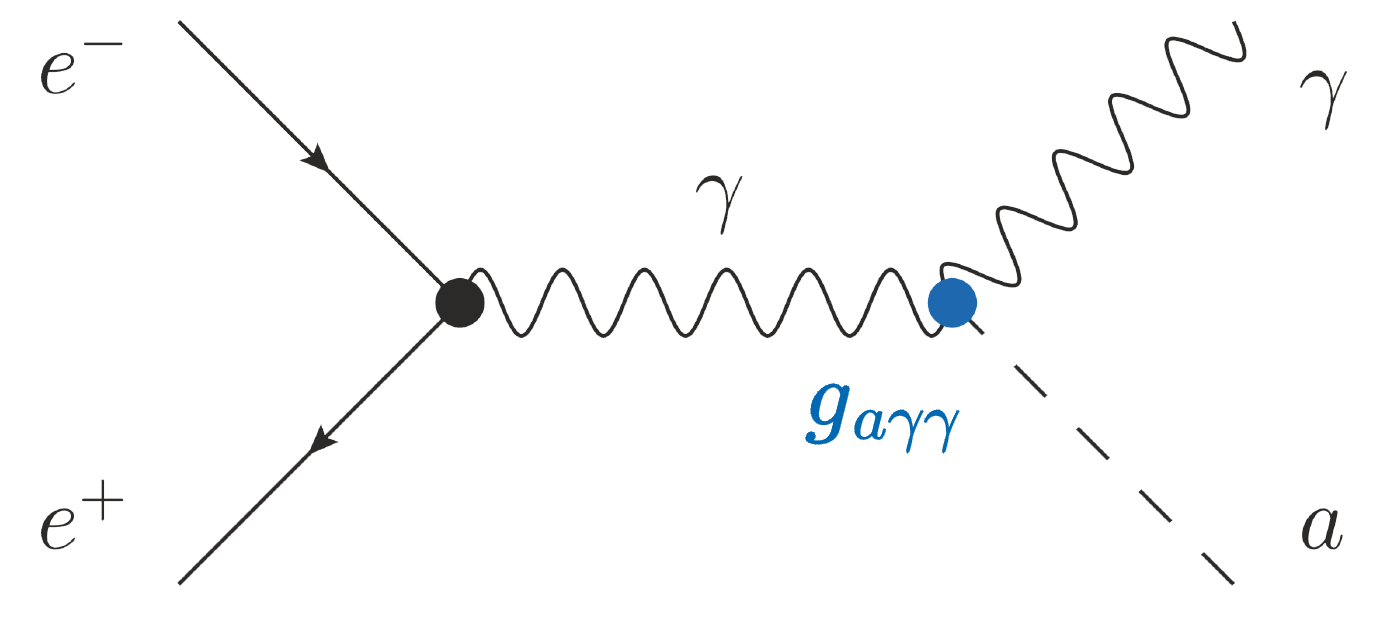}
  \put(50,-2){$(a)$}
  \end{overpic}
  \begin{overpic}[width=0.4\textwidth]{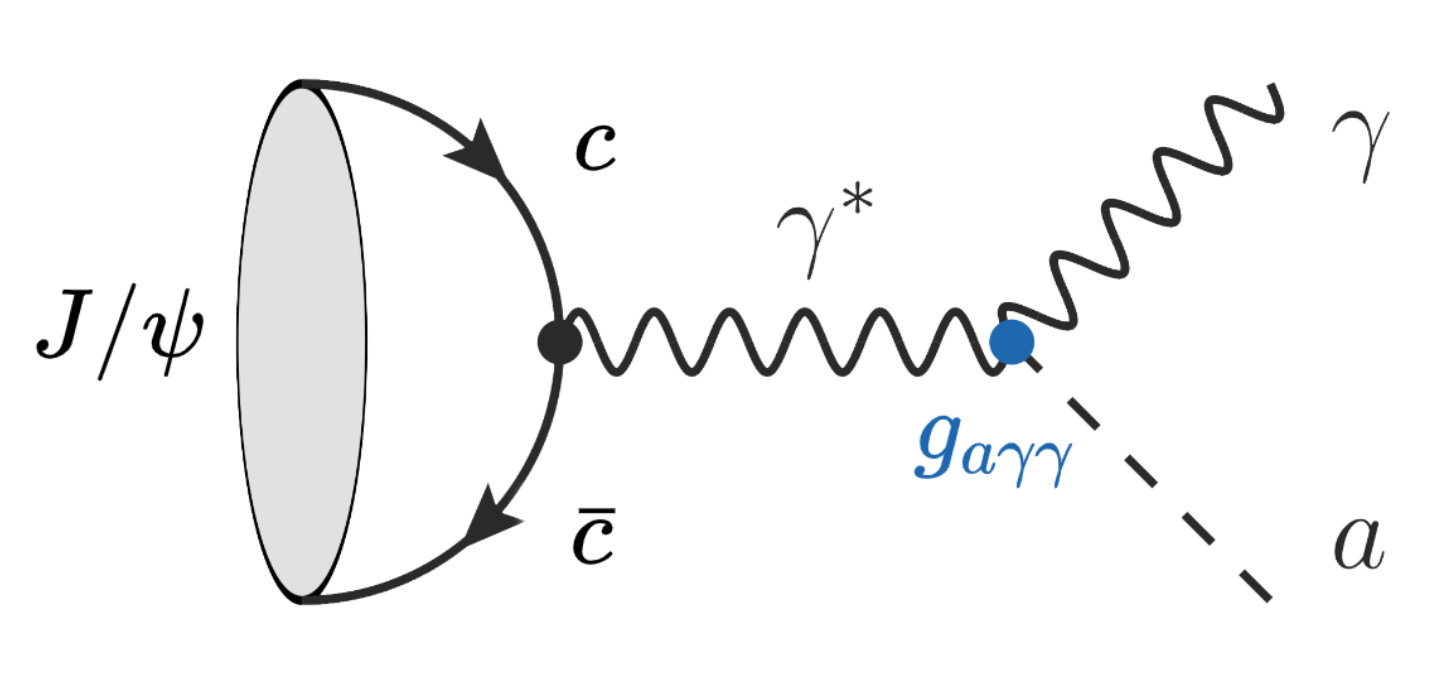}
  \put(50,-2){$(b)$}
  \end{overpic}
  }
 \caption{Feynman diagrams for (a) non-resonant ALP production and (b) resonant ALP production.}
 \label{fig:prodcution}
\end{figure}

Generally, there are two different production schemes: non-resonant production and resonant production \cite{Merlo} as shown in Figure~\ref{fig:prodcution}. The most straightforward way of producing ALPs in $e^+e^-$ facilities is via the non-resonant process $e^{+} e^{-} \rightarrow \gamma a$. Since vector quarkonia such as $J/\psi$ is a narrow resonance coupled to the electromagnetic current, it can also produce signiﬁcant resonant contributions via $e^{+} e^{-} \rightarrow J/\psi \rightarrow \gamma a$.

\section{The BESIII detector}

The BESIII detector~\cite{Ablikim:2009aa} records symmetric $e^+e^-$ collisions provided by the BEPCII storage ring~\cite{Yu:IPAC2016-TUYA01}, which operates in the center-of-mass energy range from 2.00 to 4.95 GeV, with a peak luminosity of $1\times10^{33}$~cm$^{-2}$s$^{-1}$ achieved at $\sqrt{s}$ = 3.77 GeV. BESIII has collected large data samples in this energy region~\cite{Ablikim:2019hff}. The cylindrical core of the BESIII detector covers 93\% of the full solid angle and consists of a helium-based multilayer drift chamber~(MDC), a plastic scintillator time-of-flight system~(TOF), and a CsI(Tl) electromagnetic calorimeter~(EMC), which are all enclosed in a superconducting solenoidal magnet providing a 1.0~T magnetic field. The solenoid is supported by an octagonal flux-return yoke with resistive plate counter muon identification modules interleaved with steel. 
%The acceptance of charged particles and photons is 93\% over $4\pi$ solid angle. 
The charged-particle momentum resolution at $1~{\rm GeV}/c$ is $0.5\%$, and the ${\rm d}E/{\rm d}x$ resolution is $6\%$ for electrons from Bhabha scattering. The EMC measures photon energies with a resolution of $2.5\%$ ($5\%$) at $1$~GeV in the barrel (end cap) region. The time resolution in the TOF barrel region is 68~ps, while that in the end cap region is 110~ps. The end cap TOF system was upgraded in 2015 using multigap resistive plate chamber technology, providing a time resolution of
60~ps~\cite{etof}.

\section{Search strategy}

In this article, a search for ALP resonant production in $J/\psi$ radiative decays at BESIII  via $J/\psi\rightarrow\gamma a$ is presented \cite{ALP}. The search range is $0.165\leq m_a\leq 2.84$ GeV/$c^2$ and the ALP decays to two photons with a decay width $\Gamma_{a}=g_{a \gamma \gamma}^{2} m_{a}^{3}/64 \pi$. In the search region, the decay length and decay width of the ALP are negligible compared to the detector resolution.
2.7 billion $\psi(3686)$ events with $\psi(3686) \rightarrow \pi^{+} \pi^{-} J / \psi$ decay are exploited to access the $J/\psi$ events. By using $\psi(3686)$ events, the pollution from non-resonant ALP production $e^+e^-\rightarrow\gamma a$ and large QED background could be avoided.

\begin{figure}[hp]
\centering
\includegraphics[width=0.6\textwidth]{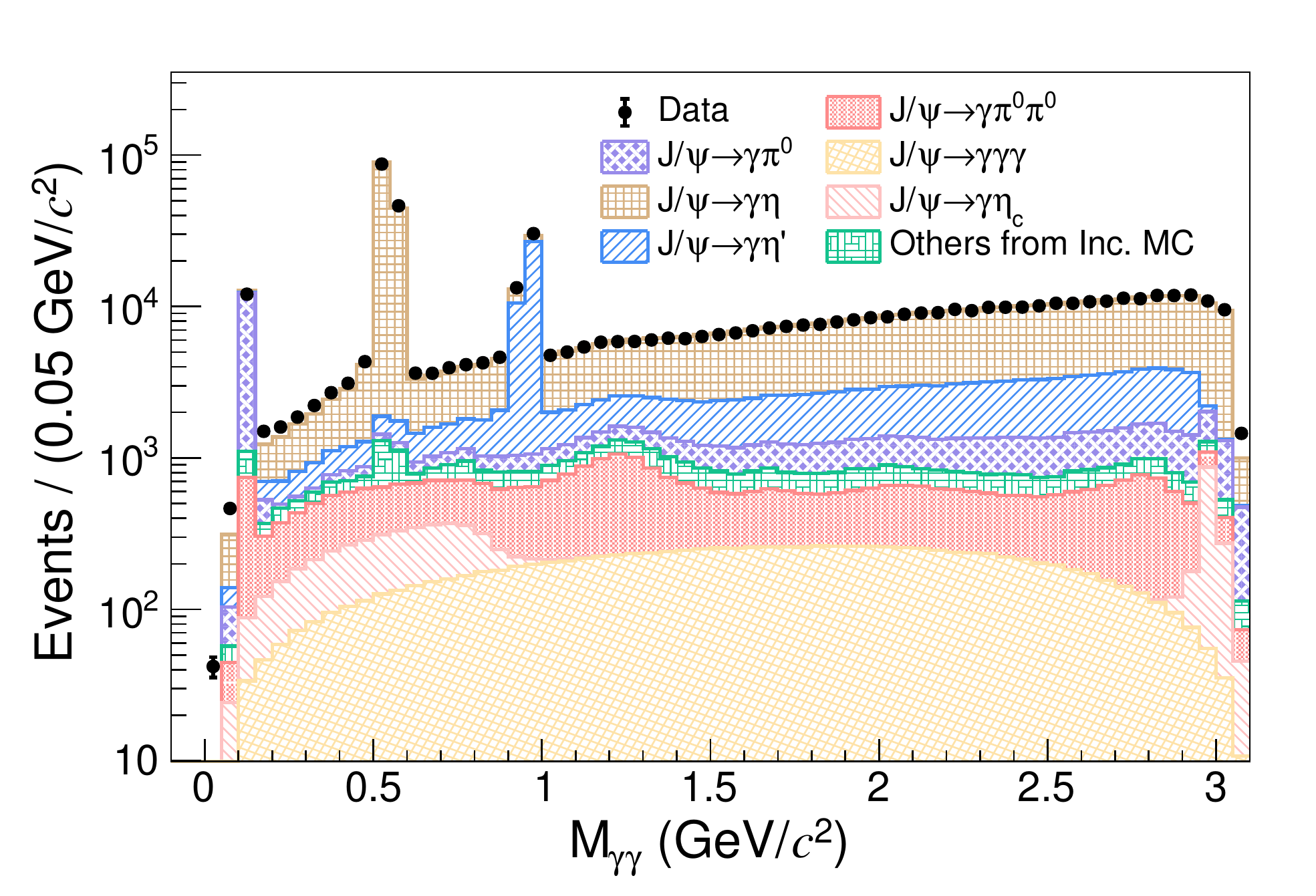}
\caption{The diphoton invariant mass distributions for data and the MC simulation backgrounds, which are normalized to the data luminosity.}
\label{fig:mgg}
\end{figure}

The two-photon invariant mass $M_{\gamma\gamma}$ distribution of the survived events after the event selection is shown in Fig.~\ref{fig:mgg}, where data and the MC simulation are well consistent. There are three entries per event from all possible combinations of the three selected photons. According to the studies of MC simulation, the background is dominated by peaking contributions from $J / \psi \rightarrow \gamma \pi^{0}$, $\gamma\eta$ and $\gamma\eta'$. The mass intervals of $0.10<m_a< 0.165$ GeV/$c^2$, $0.46<m_a< 0.60$ GeV/$c^2$ and $0.90 <m_a< 1.01$ GeV/$c^2$ are excluded due to the peaking backgrounds.

A series of unbinned extended maximum-likelihood fits are performed to the $M_{\gamma\gamma}$ distribution to determine the signal yields with 674 different ALP mass hypotheses in the mass range of $0.165\leq m_a\leq 2.84$ GeV/$c^2$.
The search step is about half the signal resolution and fit intervals depend on the specific mass hypothesis. For a specific fit, the likelihood function is a combination of signal, non-peaking background, and peaking components of the $\pi^0$, $\eta$ and $\eta'$.

\section{Results}

\begin{figure}[hp]
\centering
\includegraphics[width=0.6\textwidth]{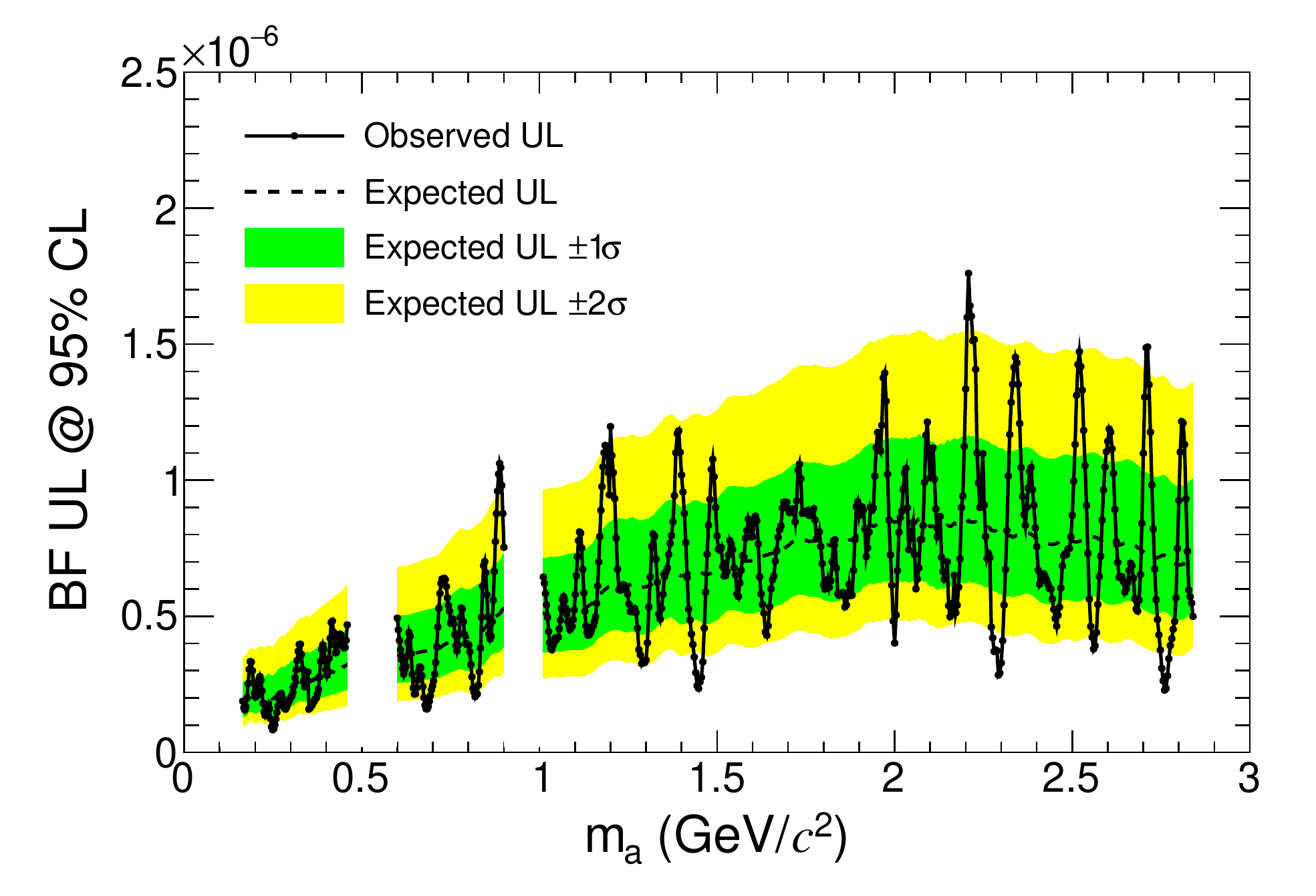}
\caption{Expected and observed upper limits at the 95\% CL on $\mathcal{B}(J/\psi\rightarrow\gamma a)$. The black curve is for data, the black dashed curve represents the expected values and the green (yellow) band represents the $\pm1\sigma$ ($\pm2\sigma$) region.}
\label{fig:UL}
\end{figure}

No significant ALP signal is observed in the search region, and the largest local significance of $2.6\sigma$ is observed near $m_a=2.208$ GeV/$c^2$, consistent with the null hypothesis. The 95\% confidence level (CL) upper limits on $\mathcal{B}(J/\psi\rightarrow\gamma a)$ as a function of $m_a$ is computed with a one-sided frequentist profile-likelihood method \cite{CLs}. There's a systematic uncertainty of 4.4\% on signal efficiency mainly from MDC tracking and photon reconstruction, and other uncertainties are considered by performing a spurious signal test and alternative fits. The expected and observed upper limits at the 95\% CL on $\mathcal{B}(J/\psi\rightarrow\gamma a)$ are shown in Fig.~\ref{fig:UL}. The observed limits range from $8.3\times10^{-8}$ to $1.8\times10^{-6}$ in the search region. 

\begin{figure}[hp]
\centering
\includegraphics[width=0.6\textwidth]{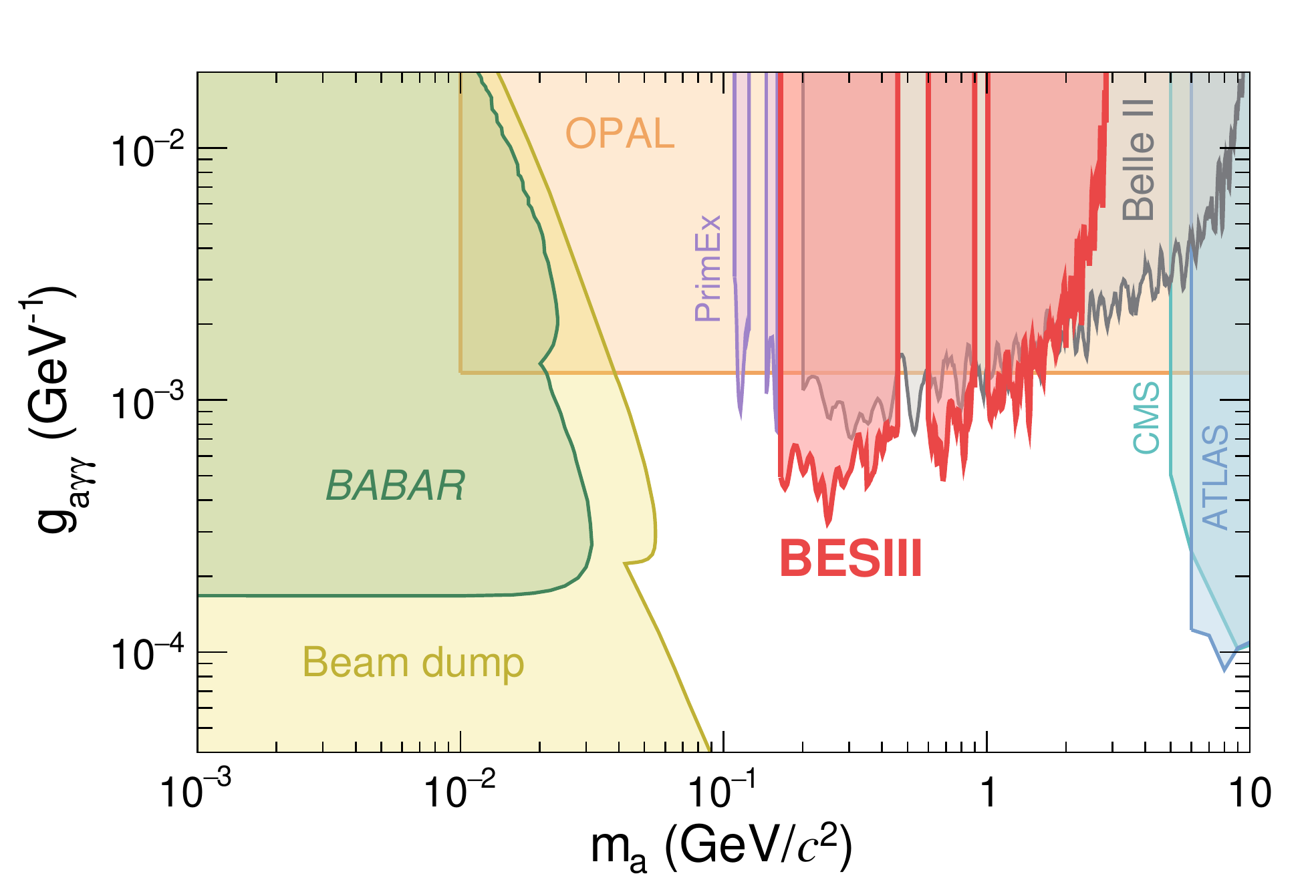}
\caption{Exclusion limits at the 95\% CL in the ALP-photon coupling $g_{a\gamma\gamma}$ versus ALP mass $m_a$ plane obtained from this analysis. All measurements assume a 100\% ALP decay branching fraction into photons.}
\label{fig:coupling}
\end{figure}

The branching fraction limit is converted to the coupling limit using \cite{Merlo}
\begin{equation}
g_{a \gamma \gamma}=\sqrt{\frac{\mathcal{B}(J / \psi \rightarrow \gamma a)}{\mathcal{B}(J / \psi \rightarrow e^+ e^-)}(1-\frac{m_{a}^{2}}{m_{J / \psi}^{2}})^{-3}\frac{32 \pi \textbf{\textbf{}}\alpha_{\mathrm{em}}}{m_{J / \psi}^{2}}},
\end{equation} 
where $\mathcal{B}(J/\psi \rightarrow e^+ e^-) = (5.971\pm0.032)\%$ is from experimental measurement \cite{pdg} and $\alpha_{\mathrm{em}}$ is the electromagnetic coupling. An additional 0.5\% uncertainty due to $\mathcal{B}(J/\psi \rightarrow e^+ e^-)$ is included when converting $\mathcal{B}(J/\psi\rightarrow\gamma a)$ to $g_{a\gamma\gamma}$. The exclusion limits in the ALP-photon coupling $g_{a\gamma\gamma}$ versus ALP mass $m_a$ plane obtained from this analysis are shown in Fig. \ref{fig:coupling}, together with the constraints of other experiments. The limits exclude the region in the ALP-photon coupling range $g_{a\gamma\gamma}>3\times10^{-4}$ for ALP mass $m_a$ around $0.25$ GeV/$c^2$, with an improvement by a factor of 2-3 over the previous Belle II measurement \cite{Belle}. In addition, the constraints on the ALP-photon coupling are the most stringent to date for $0.165\leq m_a\leq 1.468$ GeV/$c^2$.

\section{Summary}
Based on a data sample of 2.7 billion $\psi(3686)$ events collected by the BESIII detector, the resonant ALP production via $J/\psi$ radiative decays is searched using $\psi(3686)\rightarrow\pi^+\pi^- J/\psi$ process. No significant ALP signal is observed and the 95\% CL upper limits on the branching fraction of $J/\psi\rightarrow\gamma a$ and the ALP-photon coupling $g_{a\gamma\gamma}$ are set. The observed limits on $\mathcal{B}(J/\psi\rightarrow\gamma a)$ range from $8.3\times10^{-8}$ to $1.8\times10^{-6}$ in the ALP mass region of $0.165\leq m_a\leq 2.84$ GeV/$c^2$ and the exclusion limits on the ALP-photon coupling are the most stringent to date for $0.165\leq m_a\leq 1.468$ GeV/$c^2$.

\section*{Acknowledgments}

We would like to thank for the strong support from the staff of BEPCII and the IHEP computing center,
and also thank our BESIII collaborators for contributing to this proceeding.


\section*{References}

\begin{thebibliography}{99}
\bibitem{axion_1}
R. D. Peccei and H. R. Quinn,
Phys. Rev. Lett. \textbf{38}, 1440 (1977).

\bibitem{axion_2}
R. D. Peccei and H. R. Quinn,
Phys. Rev. D \textbf{16}, 1791 (1977).

\bibitem{hierarchy}
P. W. Graham, D. E. Kaplan and S. Rajendran, 
Phys. Rev. Lett. \textbf{115}, 221801 (2015).

\bibitem{Higgs}
G. C. Branco {\it et al.},
Phys. Rept. \textbf{516}, 1 (2012).

\bibitem{dark_1}
M. Freytsis and Z. Ligeti,
Phys. Rev. D \textbf{83}, 115009 (2011).

\bibitem{Merlo}
L. Merlo {\it et al.},
JHEP \textbf{06}, 091 (2019).

\bibitem{Ablikim:2009aa}
M.~Ablikim {\it et al.} (BESIII Collaboration),
%``Design and Construction of the BESIII Detector,''
Nucl.\ Instrum.\ Meth.\ A {\bf 614}, 345 (2010).

\bibitem{Yu:IPAC2016-TUYA01}
C.~H.~Yu {\it et al.},
%``BEPCII Performance and Beam Dynamics Studies on Luminosity,''
Proceedings of IPAC2016, Busan, Korea, 2016,
doi:10.18429/JACoW-IPAC2016-TUYA01.

\bibitem{Ablikim:2019hff}
M.~Ablikim {\it et al.} (BESIII Collaboration),
%``White Paper on the Future Physics Programme of BESIII,''
Chin. Phys. C {\bf 44}, 040001 (2020).

\bibitem{etof}
X.~Li {\it et al.}, Radiat. Detect. Technol. Methods {\bf 1}, 13 (2017);
Y.~X.~Guo {\it et al.}, Radiat. Detect. Technol. Methods {\bf 1}, 15 (2017);
P.~Cao {\it et al.}, Nucl.\ Instrum.\ Meth.\ A {\bf 953}, 163053 (2020).

\bibitem{ALP}
M.~Ablikim {\it et al.} (BESIII Collaboration),
Phys. Lett. B {\bf 838}, 137698 (2023).

\bibitem{CLs}
G. Cowan {\it et al.},
Eur. Phys. J. C \textbf{71}, 1554 (2011);
[Erratum: Eur. Phys. J. C \textbf{73}, 2501 (2013)].

\bibitem{pdg}
R. L. Workman {\it et al.} (Particle Data Group), Prog. Theor. Exp. Phys. {\bf 2022}, 083C01 (2022)

\bibitem{Belle}
F. Abudinén {\it et al.} (Belle II Collaboration), 
Phys. Rev. Lett. \textbf{125}, 161806 (2020).


\end{thebibliography}
\end{document}